\documentclass[11pt,aps,superscriptaddress,preprintnumbers,amsmath,amssymb,nofootinbib]{revtex4}
\usepackage{epsfig}  
\usepackage{graphicx}
\usepackage{hyperref}
\usepackage{color}
\usepackage{float}
\usepackage{amsfonts}
\usepackage{amsmath}
\usepackage{slashed}
\usepackage{float}

\large

\begin{document} 

\title{Impact of $b \to s \ell \ell$ anomalies on rare charm  decays in  non-universal  $Z'$ models}

\author{Ashutosh Kumar Alok}
\email{akalok@iitj.ac.in}
\affiliation{Indian Institute of Technology Jodhpur, Jodhpur 342037, India}

\author{Neetu Raj Singh Chundawat}
\email{chundawat.1@iitj.ac.in}
\affiliation{Indian Institute of Technology Jodhpur, Jodhpur 342037, India}

\author{Dinesh Kumar}
\email{dinesh@uniraj.ac.in}
\affiliation{Department of Physics, University of Rajasthan, Jaipur 302004, India}

\begin{abstract}
In this work, we study the impact of $b \to s \ell \ell$, $B_s - \bar{B_s}$ mixing and neutrino trident measurements on observables in decays induced by $c \to u $  transition in the context of a non-universal $Z'$ model which generates $C^{\rm NP}_{9} <0$ and $C^{\rm NP}_9  = - \,C^{\rm NP}_{10} $ new physics scenarios at the tree level. We inspect the effects on $D^0 \to \pi^0 \nu \bar{\nu}$, $D^+ \to \pi^+ \nu \bar{\nu}$ and $B_c \to B^+ \nu \bar{\nu} $ decays which are induced by the quark level transition $c \to u \nu \bar{\nu}$. The fact that the branching ratios of these decays are negligible in the standard model (SM) and the long distance effects are relatively smaller in comparison to their charged dileptons counterparts, they are considered to provide genuine null-tests of SM. Therefore the observation of these modes at the level of current as well as planned experimental sensitivities would imply unambiguous signature of new physics.  Using the constraints on $Z'$ couplings coming from a combined fit to $b \to s \ell \ell$, $\Delta M_s$  and neutrino trident data, we find that any meaningful enhancement over the SM value is ruled out in the considered framework. The same is true for $D - \bar{D}$  mixing observable $\Delta M_D$ along with  $D^0 \to \mu^+ \mu^-$ and $D^+ \to \pi^+ \mu^+ \mu^-$ decay modes which are induced through $c \to u \mu^+ \mu^-$ transition.
\end{abstract}

\maketitle 

\newpage

\section{Introduction} 
The currently running experiments at the LHC have provided several engrossing indirect  hints  of physics beyond the standard model (SM) of electroweak interactions. Many such signatures of new physics are in observables related to the decays induced by the quark level transition  $b\rightarrow s\, \ell^+\,\ell^-$ $(l=e,\,\mu)$ \cite{London:2021lfn}. The most interesting discrepancies are related to lepton flavour universality (LFU) violation in decays $B \to (K, \, K^*) \, l^+\,l^-$ as within the SM, the coupling of leptons of gauge bosons are flavour independent. The measurements of LFU testing ratios $R_K \equiv \Gamma(B^+ \to K^+ \mu^+ \mu^-)/\Gamma(B^+ \to K^+ e^+ e^-)$ \cite{LHCb:2021trn} and $R_{K^*} \equiv \Gamma(B^0 \to K^{*0} \mu^+ \mu^-)/\Gamma(B^0 \to K^{*0} e^+ e^-)$  \cite{rkstar} disagree with the SM prediction which is very close to unity  \cite{Hiller:2003js,Bordone:2016gaq,Isidori:2020acz}, with small corrections due to the muon-electron mass difference,  at the level of 3.1 and 2.5$\sigma$, respectively. Very recently such ratios were measured in $B^0\to K^0_S l^+ l^-$ and $B^+\to K^{*+}\ell^+\ell^-$ decay modes by the LHCb collaboration using data set corresponding to an integrated luminosity  of  9 $\rm fb^{-1}$ \cite{LHCb:2021lvy}. These measurements have relatively large uncertainties and
 they are  consistent with the SM value at 1.5 and 1.4$\sigma$, respectively. These discrepancies can be attributed to new physics in muon or/and electron sector. 
 
 There is another class of discrepancies that can be imputed to new physics only in the muon sector. These are in angular observables constructed from kinematical distribution of differential decay widths of $B \to K^* \,  \mu^+\,\mu^-$  and $B_s \to \phi\, \mu^+\,\mu^-$ decays. A 	series of measurements from different experiments at the LHC along with the Belle experiment  have reported deviations from their SM predictions of the angular observable  $P'_5$ in $B \to K^* \, \mu^+\,\mu^-$ decay at the level of 3$\sigma$ \cite{Kstarlhcb1,Kstarlhcb2, sm-angular}. In addition, the measured value of the branching ratio of $B_s \to \phi\, \mu^+\,\mu^-$ decay shows tension with the SM prediction at the level of 3.5$\sigma$ \cite{bsphilhc2} .
 
 The piling up of these disparateness with the SM can be considered as signals of new physics. The Lorentz structure of this possible new physics  can be determined by a model-independent analysis within the framework of effective field theory (EFT) where the new physics effects  are incorporated by adding new operators $O_i$ to the SM effective Hamiltonian. The contributions of heavy new particles are integrated out and decoded in the Wilson coefficients (WCs) $C_i$ of the operators $O_i$. A global analyses of all relevant data in  $b\rightarrow s\, \ell^+\,\ell^-$  sector have been performed by several groups, see for e.g. \cite{Descotes-Genon:2013wba, Altmannshofer:2013foa,Alok:2019ufo,Altmannshofer:2021qrr,Carvunis:2021jga,Alguero:2021anc,Geng:2021nhg,Hurth:2021nsi,Angelescu:2021lln}. These analyses differ mainly in the treatment of hadronic uncertainties and the statistical approach. Irrespective of the adopted methodology in these analyses,  new physics scenarios with non-zero muonic WCs corresponding to vector/axial-vector operators are statistically favoured.

The global analyses of  $b\rightarrow s\, \ell^+\,\ell^-$  data suggest various new physics solutions. Some of these solutions are preferred over the SM with a very high significance.  For e.g., considering one new physics operator  or two related operators at a time and assuming new physics only in the muon sector,  the $O_9\, \equiv  (\overline{s} \gamma^{\mu} P_L b)(\overline{\mu} \gamma_{\mu}\mu)$ operator as well as a combination of $O_9$  and $O_{10}\, \equiv  (\overline{s} \gamma^{\mu} P_L b)(\overline{\mu} \gamma_{\mu} \gamma_5  \mu)$  with $C_9^{NP} = - \, C_{10}^{NP}$ can account for all  $b\rightarrow s\, \ell^+\,\ell^-$  data.  These model independent solutions can be realized in  several new physics models. One such simplified model is $Z'$ with non-universal couplings which generates $b\rightarrow s\, \mu^+\,\mu^-$  decay mode at the tree level, see for e.g. \cite{Chang:2010zy,Buras:2013qja,Chang:2013hba,Altmannshofer:2014cfa,Crivellin:2015lwa,Sierra:2015fma,Crivellin:2015era,Allanach:2015gkd,Boucenna:2016wpr,Boucenna:2016qad,Datta:2017pfz, Darme:2018hqg,Calibbi:2019lvs,Alok:2021ydy,Mohapatra:2021ynn,Crivellin:2020oup}. Various versions of these models are widely studied in the context of $B$-anomalies. However, the $Z'$ boson can also contribute to other sectors. The muon coupling to $Z'$  is constrained by $b \to s$ and neutrino trident data and hence it is possible to look for implications of such couplings in other sector as  well. Such a correlation  in $b \to d$ sector was studied in \cite{Alok:2019xub}.
 
 In this work we study implications of current $b \to s$ and neutrino trident data on several observables in the charm  sector in the context of $Z'$ models. In particular we study decays induced by the quark level transition $c \to u \nu \bar{\nu}$ and $c \to u \mu^+ \mu^-$.  These rare decays provide unique opportunity for probing new physics in the up quark sector. The SM symmetries are utilized to define null-test observables with small theoretical uncertainties in these semi-leptonic decays  which are usually dominated by vector resonances \cite{Gisbert:2020vjx}. Therefore it would be interesting to investigate the impact of possible new physics in $b \to s$ sector on charm decays which can play complementary role in  understanding the nature and origin of new physics apparent in the down sector. The theoretical potential embedded in the charm sector can be complemented and utilized by the  several ongoing and planned experimental facilities such as LHCb \cite{LHCb:2008vvz}, Belle II \cite{Belle-II:2018jsg}, BES III \cite{BESIII:2020nme}, the FCC-ee \cite{FCC:2018byv} and Super Charm-Tau factory \cite{Charm-TauFactory:2013cnj}.
 
The paper is organized as follows. In section~\ref{sec:II}, we introduce the $Z'$ model considered in this work along with its contributions to the effective Hamiltonian of $b \to s$ and $c \to u$ processes. In the next section we discuss observables which provide useful constraints on $Z'$ couplings relevant for the  $b \to s$ sector. We also provide the fit results. Using the constraints obtained on new physics couplings, in section~\ref{sec:IV}, we provide predictions for several observables in the charm sector. We also show correlations between potential observables in charm and $b \to s \mu^+ \mu^-$ sectors.  Finally, the conclusions of this work are illustrated in section~\ref{sec:VI}.

 \section{The non-universal $Z'$ model}
\label{sec:II}
  We consider a non-universal $Z'$ model where the $Z'$ boson is associated with an additional $U(1)'$ symmetry.  In this model, the $b\rightarrow s\, \mu^+\,\mu^-$ transition is generated at the tree level. The $Z'$ boson  
 couples to both left and right-handed muons but not to leptons of other generations. Further, the couplings to both left and right-handed quarks are allowed. However, in order to avoid contribution of new chirality flipped operators to flavour changing neutral current (FCNC) decays, the couplings to right-handed quarks are assumed to be  flavour-diagonal \cite{Barger:2009eq, Barger:2009qs}. As we are interested in $b\rightarrow s$ processes, the change in the Lagrangian density due to the addition of this heavy $Z'$ boson can be written as
 \begin{equation}
 \Delta \mathcal{L}_{Z'} = J^{\alpha}Z'_{\alpha}\;,
 \end{equation}
where
\begin{eqnarray}
  J^{\alpha} &\supset & g^{\mu\mu}_L\, \bar{L}\gamma^{\alpha}P_L L + g^{\mu\mu}_R\, \bar{\mu}\gamma^{\alpha}P_R\, \mu 
+ g^{bs}_L\, \bar{Q}_2 \gamma^{\alpha}P_L Q_3
  + h.c. \,,
\label{eq:Jalpha}
\end{eqnarray}
where $g_{L(R)}^{\mu\mu}$ are the left-handed (right-handed) couplings of the $Z'$ boson to muons, and $g_L^{bs}$ to quarks. The right handed quark coupling, $g_R^{bs}$,  cannot contribute to the $b \to s$ processes as these couplings are assumed to be flavour diagonal. Further,   $P_{L(R)} = (1\mp \gamma_5)/2$, $Q_{i}$ is
the $i^{th}$ generation of quark doublet, and $L = (\nu_{\mu}, \mu)^T$ is the second generation doublet. 
 
After integrating out  the heavy $Z'$, we get  the  effective four-fermion Hamiltonian which apart from contributing to $b\rightarrow s\, \mu^+\,\mu^-$ transition, also induces  $b\rightarrow s\, \nu\, \bar{\nu}$ and $c\rightarrow u$  transitions. The relevant terms in the effective Hamiltonian is given by
\begin{eqnarray}
  \mathcal{H}_{\rm eff}^{Z'} &=& \frac{1}{2M^2_{Z'}}J_{\alpha}J^{\alpha} \supset  \frac{g^{bs}_L}{M^2_{Z'}} \left(\bar{s}\gamma^{\alpha}P_L b\right)
  \left[\bar{\mu}\gamma_{\alpha}\left(g^{\mu\mu}_L P_L 
   + g^{\mu\mu}_R P_R\right)\mu \right]  \nonumber \\
 & & 
   +  \frac{\left(g^{bs}_L\right)^2}{2M^2_{Z'}}\left(\bar{s}\gamma^{\alpha}P_L b\right)\left(\bar{s}\gamma_{\alpha}P_L b\right)\nonumber \\
 & & 
  + \frac{g^{\mu\mu}_L}{M^2_{Z'}} \left(\bar{\nu}_{\mu}\gamma_{\alpha}P_L\nu_{\mu}\right) \left[\bar{\mu}\gamma^{\alpha}\left(g^{\mu\mu}_L P_L  + g^{\mu\mu}_R P_R\right)\mu\right]
   \nonumber\\
  & & + \frac{g^{bs}_L \, g^{\mu\mu}_L}{M^2_{Z'}}\left(\bar{s}\gamma^{\alpha}P_L b\right)\left(\bar{\nu}_{\mu}\gamma_{\alpha}P_L \nu_{\mu}\right)
  \nonumber\\
  & & + \frac{h^{cu}_L}{M^2_{Z'}} \left(\bar{u}\gamma^{\alpha}P_L c\right)
  \left[\bar{\mu}\gamma_{\alpha}\left(g^{\mu\mu}_L P_L 
   + g^{\mu\mu}_R P_R\right)\mu \right]
   \nonumber \\
 & & + \frac{\left(h^{cu}_L\right)^2}{2M^2_{Z'}}\left(\bar{u}\gamma^{\alpha}P_L c\right)\left(\bar{u}\gamma_{\alpha}P_L c\right)
   \nonumber\\
   & & + \frac{h^{cu}_L \, g^{\mu\mu}_L}{M^2_{Z'}}\left(\bar{u}\gamma^{\alpha}P_L c\right)\left(\bar{\nu}_{\mu}\gamma_{\alpha}P_L \nu_{\mu}\right)\,,
 \label{Leff}
 \end{eqnarray}
where $h^{cu}_L = g_L^{bs} V_{us} V_{cb}^{*}$. 
The fact that the down-type quarks in the quark-doublets $Q_i$ are taken to be in the mass-flavour diagonal basis, $u_i \to u_j$ transitions are induced by the up-type quarks in the quark doublets. 

The first term in eq.~(\ref{Leff}) induces $b\rightarrow s  \mu^+\mu^-$ transition whereas the second term gives rise to $B_{s}$--$\bar{B}_{s}$ mixing. The third term  contributes to the neutrino trident production $\nu_{\mu} N \rightarrow \nu_{\mu} N \mu^+\mu^-$ ($N$ = nucleus). The fourth term generates $b \to s \, \nu\, \bar{\nu}$ decay whereas the remaining terms induces FCNC transitions in the up quark sector. The contribution to $c \to u \mu^+ \mu^-$ and $D^0$-$\bar{D^0}$ in this model can be calculated through fifth and sixth terms, respectively.
The seventh term corresponds to charm dineutrino decay. The product $g_L^{bs} g_{L,R}^{\mu\mu} $ and the  individual magnitude $|g_{L}^{bs}|$  are constrained by the $b \to s \mu^+ \mu^-$ data and $B_{s}$--$\bar{B}_{s}$ mixing, respectively.   The constraints on individual muon couplings $g_{L,R}^{\mu\mu}$ are obtained through the measurement of the neutrino trident production. Observables in $c \to u$  sectors depend upon specific combinations of these couplings which can be determined from  eq.~(\ref{Leff}). Thus it would be interesting to see the implications of  $b \to s \mu^+ \mu^-$,  $B_{s}$--$\bar{B}_{s}$ and neutrino trident data on possible new physics effects in  charm decays. We now discuss contributions of  $Z'$ boson to these processes. 

\subsection{$b\rightarrow s $  transitions}
In this subsection, we calculate the contributions of $Z'$ boson to $b\rightarrow s $  transitions. Here we consider $b\rightarrow s \mu^+\mu^-$ and $b \to s \nu \bar{\nu}$ decays along with $B_s-\bar{B}_s$ mixing. 

\subsubsection{$b\rightarrow s \mu^+\mu^-$ decay}

The effective Hamiltonian for  $b\rightarrow s \mu^+\mu^-$ transition in the SM is given by
\begin{align} \nonumber
  \mathcal{H}_{\rm eff}^{\rm SM} &= -\frac{ 4 G_F}{\sqrt{2}} V_{ts}^* V_{tb}
  \bigg[ \sum_{i=1}^{6}C_i {O}_i + C_8 { O}_8   + C_7\frac{e}{16 \pi^2}[\overline{q} \sigma_{\mu \nu}
      (m_s P_L + m_b P_R)b] F^{\mu \nu}   \nonumber\\
     & + C^{\rm SM}_9 \frac{\alpha_{\rm em}}{4 \pi}
    (\overline{s} \gamma^{\mu} P_L b)(\overline{\mu} \gamma_{\mu} \mu)  
    + C^{\rm SM}_{10} \frac{\alpha_{\rm em}}{4 \pi}
    (\overline{s} \gamma^{\mu} P_L b)(\overline{\mu} \gamma_{\mu} \gamma_{5} \mu)
    \bigg] \;.
\end{align}
Here $G_F$ is the Fermi constant and $V_{ij}$ are the Cabibbo-Kobayashi-Maskawa
(CKM) matrix elements.
The short distance contributions are encoded in the Wilson coefficients (WC) $C_i$ of the four-fermi operators ${O}_i$ where the scale-dependence is implicit, i.e. $C_i \equiv C_i(\mu)$
and ${O}_i \equiv {O}_i(\mu)$. 
The contributions of operators ${O}_i$ ($i=1,...,6,8$) to $b\rightarrow s \mu^+\mu^-$ are included through the modifications  $C_{7,9}(\mu)$ $\rightarrow$
$ C_{7,9}^{\mathrm{eff}}(\mu,q^2)$. Here $q^2$ is the invariant mass-squared of the final state muon pair.
The $Z'$ boson contributes to $b\rightarrow s \mu^+\mu^-$ decay mode through the first term in 
eq.~(\ref{Leff}). This contribution modifies the SM WCs $C^{\rm SM}_{9,10}$  as $C_{9,10} \rightarrow C^{\rm SM}_{9,10} + C^{\rm NP}_{9,10}$. The new contributions $C^{\rm NP}_{9,10}$ are given as
 \begin{eqnarray}
  C^{\rm NP}_9 &=& -\frac{\pi}{\sqrt{2}G_F \, \alpha_{\rm em} V_{tb}V^*_{ts}} \frac{g_L^{bs}(g_L^{\mu\mu}+g_R^{\mu\mu})}{M^2_{Z'}}\,, \nonumber\\
  C^{\rm NP}_{10} &=& \frac{\pi}{\sqrt{2}G_F\,\alpha_{\rm em} V_{tb}V^*_{ts}} \frac{g_L^{bs}(g_L^{\mu\mu}- g_R^{\mu\mu})}{M^2_{Z'}}\,.
  \label{bqllNP}
\end{eqnarray}
It is well known that the new physics scenario $C^{\rm NP}_{9} <0$ as well as $C^{\rm NP}_9  = - \,C^{\rm NP}_{10} $ provide a good fit to all  $b\rightarrow s \mu^+\mu^-$ data. It is evident from eq.~(\ref{bqllNP}) that these  one-dimensional (1D) new physics solutions can be generated 
by substituting $g^{\mu\mu}_L = g^{\mu\mu}_R$ and $g^{\mu\mu}_R = 0$, respectively. In this work, we consider both scenarios. We denote  $g^{\mu\mu}_L = g^{\mu\mu}_R$  and $g^{\mu\mu}_R = 0$ scenarios as $Z_1$ and $Z_2$, respectively. 

\subsubsection{$B_s-\bar{B}_s$ mixing}

The $B_s$ mixing is generated in the SM through the box diagrams.  The dominant contribution comes from the virtual top quark in the loop. The mass difference of the two mass eigenstates, $\Delta M_s \equiv M_H^s - M_L^s$, is two times $M^s_{12}$, the dispersive part of the box diagrams responsible for the mixing. Within SM, $M^s_{12}$ is given by
\begin{equation}
M^{s,\rm SM}_{12}= \frac{G_F^2 M_W^2}{12 \pi^2} \left(V_{tb}V^*_{ts}\right)^2 M_{B_s}f_{B_s}^2 \widehat{B}_{B_s} \eta_B S_0(x_t)\,,
\end{equation}
where $x_t \equiv m_t^2/M_W^2$, $S_0(x_t)$ is the Inami-Lim function \cite{Inami:1980fz} and $\eta_B \approx 0.84$ encodes perturbative QCD corrections at two loop \cite{Buras:1990fn}. The $Z'$ boson generates $B_s-\bar{B}_s$ mixing at the tree-level through the second term in eq.~(\ref{Leff}) which is the same operator as in the SM. The combined contribution is written as \cite{DiLuzio:2019jyq}
\begin{equation}
\frac{\Delta M_s^{\rm SM+Z'}}{\Delta M_s^{\rm SM}} \approx  \left|1+ (5 \times 10^{3}) (g^{bs}_L)^2 \right|\,,
\label{eq:Bsmix}
\end{equation}
for $M_{Z'}$ = 1 TeV.

\subsubsection{$b \to s \nu \bar{\nu}$ decay}
 Like $b\rightarrow s l^+ l^-$ transition, $b \to s \nu \bar{\nu}$ occurs via one-loop electroweak penguin or box diagrams in the SM. However $b \to s \nu \bar{\nu}$ decay rates are predicted with relatively smaller theoretical uncertainties in comparison to the corresponding 
 $b\rightarrow s l^+ l^-$ decay modes. This is mainly due to the absence of the long-distance 
 electromagnetic interactions. The effective Hamiltonian for $b\to s \nu \bar{\nu}$ transition  is given by \cite{Buras:2014fpa}
\begin{equation}
H_{\rm eff} = - \frac{\sqrt{2} G_F \, \alpha_{\rm em} }{\pi} V_{tb} V_{ts}^* \sum_\ell
C_L^\ell (\bar s \gamma_{\mu} P_L b) (\bar \nu_\ell \gamma^{\mu}
P_L\nu_\ell) ~.
\end{equation}
Here $C_L^\ell = C_L^{\rm SM} + C_\nu^{\ell\ell}({\rm NP})$. 
The SM WC is $C_L^{\rm SM} = - X_t/s_W^2$, where $s_W \equiv \sin\theta_W$ and $X_t = 1.469 \pm 0.017 $. The NP contribution $C_\nu^{\mu\mu}({\rm NP})$ in the $Z'$ model is  given by
  \begin{equation}
  C_\nu^{\mu\mu}({\rm NP}) =  -\frac{\pi}{\sqrt{2}G_F\,\alpha_{\rm em} V_{tb}V^*_{ts}} \frac{g_L^{bs} g_L^{\mu\mu}}{M^2_{Z'}}\,.
  \end{equation}

\subsection{Neutrino trident production}
Due to $SU(2)_L$ invariance, the $Z'$ also couples to the left-handed neutrinos. This leads to the neutrino trident production, $\nu_{\mu} N \rightarrow \nu_{\mu} N \mu^+\mu^-$. The the third term in eq.~(\ref{Leff}) modifies the cross section $\sigma$ for neutrino trident production as \cite{Alok:2017jgr}
\begin{eqnarray}
R_\nu =  \frac{\sigma}{\sigma_{\rm SM}} &=& \frac{1}{1+(1+4s^2_W)^2}\Bigg[\left(1+ \frac{v^2g^{\mu\mu}_L(g^{\mu\mu}_L-g^{\mu\mu}_R)}{M^2_{Z'}}\right)^2 \nonumber\\
&&
+ \left(1+4s^2_W+\frac{v^2g^{\mu\mu}_L(g^{\mu\mu}_L+g^{\mu\mu}_R)}{M^2_{Z'}}\right)^2\Bigg],
\label{trident}
\end{eqnarray}
where $v=246$ GeV and $s_W = \sin\,\theta_W$.

\subsection{$c \to u $ transitions}
The $Z'$ contributes to $c \to u $ transition through fifth, sixth and seventh terms in eq.~(\ref{Leff}). These terms induce $c \to u \mu^+ \mu^-$ decay, $D$-$\bar{D}$ mixing and   $c \to u \nu \bar{\nu}$ transition, respectively. 

\subsubsection{$c \to u \mu^+ \mu^-$ decay}
The $c \to u \mu^+ \mu^-$  process  in the SM can be described by the following effective Hamiltonian at the $\mu_c=m_c$ scale \cite{Greub:1996wn,Burdman:2001tf,Fajfer:2002gp,Dorsner:2009cu,deBoer:2015boa,deBoer:2016dcg,Fajfer:2015mia,Bause:2020obd}
\begin{equation}
  \mathcal{H}_{\rm eff}^{\rm SM} ( c \to u \mu^+ \mu^-) =  (V_{ud} V_{cd}^*) \, \mathcal{H}^{d}
  + (V_{us} V_{cs}^*)\,  \mathcal{H}^{s} + (V_{ub} V_{cb}^*) \, \mathcal{H}^{\rm peng},
\end{equation}
  where the three contributions correspond to diagram with intermediate quarks $d,s,b$. It is customary to include the contributions of the states heavier than the charm quark in $\mathcal{H}^{\rm peng}$ as
  \begin{equation}
  \mathcal{H}^{\rm peng} = -\frac{ 4 G_F}{\sqrt{2}} \sum_{i=3,....,10} C_i^{cu} \, O_i\,.
  \end{equation}
  The following operators appearing in the above effective Hamiltonian are in particular sensitive to new physics effects
  \begin{eqnarray}
  O_7 &=& \frac{e m_c}{4 \pi^2} \left(\overline{u} \sigma_{\mu \nu}
      P_R c \right) F^{\mu \nu}\,, \\
  O_9 &=& \frac{\alpha_{\rm em}}{4 \pi}  (\overline{u} \gamma^{\mu} P_L c)(\overline{\mu} \gamma_{\mu} \mu) \,,\\
  O_{10} &=&  \frac{\alpha_{\rm em}}{4 \pi}
    (\overline{u} \gamma^{\mu} P_L c)(\overline{\mu} \gamma_{\mu} \gamma_{5} \mu).
  \end{eqnarray}
 The only non-vanishing WCs in the SM are $C_{7,9}^{cu}$. Their values are determined mainly    through the effects of QCD renormalization. The value of $C_7^{cu}(m_c)$ determined by two loop mixing with current-current operators  is $V_{ub} V_{cb}^* C_7^{cu}  = V_{us} V_{cs}^*$ $(0.007 + 0.020i) (1 \pm 0.2)$ \cite{Greub:1996wn,Ho-Kim:1999jjf}. After inclusion of  renormalization group running effects, the value of $C_9^{cu}(m_c)$ turns out to be small \cite{Fajfer:2002gp}. On the other hand, as the renormalization group running does not affect $O_{10}$, the WC  $C_{10}^{cu}(m_c)$ is negligibly small \cite{Fajfer:2005ke}. As $C_{10}^{cu} \approx 0$, the effects of $V-A$ structure of the SM  are switched off at the charm scale. This feature makes FCNC charm decays distinct from that of the corresponding $B$ and $K$ decays. 

The $Z'$ boson contributes to $c \to u \mu^+ \mu^-$ decay through the fifth term in eq.~(\ref{Leff}). This contribution modifies the SM WCs $C^{cu}_{9,10}$  as $C^{cu, \rm tot}_{9,10} \rightarrow C^{cu}_{9,10} + C^{cu,  Z'}_{9,10}$. The new contributions $C^{cu, Z'}_{9,10}$ are given by
 \begin{eqnarray}
  C^{cu, Z'}_9 &=& -\frac{\pi}{\sqrt{2}G_F\,\alpha_{\rm em} V_{ub}V^*_{cb}} \frac{h_L^{cu}(g_L^{\mu\mu}+g_R^{\mu\mu})}{M^2_{Z'}}\,, \nonumber\\
  C^{cu, Z'}_{10} &=& \frac{\pi}{\sqrt{2}G_F\,\alpha_{\rm em} V_{ub}V^*_{cb}} \frac{h_L^{cu}(g_L^{\mu\mu}- g_R^{\mu\mu})}{M^2_{Z'}}\,.
  \label{cllNP}
\end{eqnarray}
For $Z_1$ ($Z_2$) model, $g^{\mu\mu}_L = g^{\mu\mu}_R$ ($g^{\mu\mu}_R = 0$).

\subsubsection{$D$-$\bar{D}$ mixing}
In the SM, $D^0$-$\bar{D^0}$ mixing is induced by the box diagrams with  d, s and b quarks in the loop. Due to a strong GIM cancellation, 
the short-distance contribution is extremely small,  $\Delta M_D=O(10^{-4})\, \rm ps^{-1}$. Therefore the short-distance contribution cannot explain the measured value of $\Delta M_D$ which is $(0.95^{+0.41}_{-0.44}) \times 10^{-2}\, \rm ps^{-1}$ \cite{pdg}. In particular, the contribution 
due to  b-quark is highly suppressed, $O(\lambda^8)$. The fact that $D^0$-$\bar{D^0}$ mixing is dominated by the d- and s-quarks, there can be large long-distance contributions, for which there are no reliable estimates at present \cite{Petrov:2006nc,Golowich:2009ii}. As long-distance contributions are unknown, in our analysis we focus on the short-distance contributions to the  $D^0$-$\bar{D^0}$ mixing parameter $\Delta M_D$.

In $Z'$ model, $D^0$-$\bar{D^0}$ mixing is induced at the tree level and hence  expected to provide a much larger contribution in comparison to the short-distance SM contribution. $\Delta M_D$ generated by the sixth term in eq.~(\ref{Leff}) is given by
\begin{equation}
\Delta M_D = \frac{f_D^2\, m_D\, B_D\, r(m_c,M_{Z'})}{3M^2_{Z'}} |(h^{cu}_L)^2| \,,
\end{equation}
where $f_D = 212.0 \pm 0.7$ MeV \cite{FlavourLatticeAveragingGroup:2019iem}, $B_D = 0.757 \pm 0.027 \pm 0.004$ \cite{Carrasco:2015pra} and the RG factor $r(m_c,M_{Z'}) = 0.72$ for $M_{Z'} = 1$ TeV \cite{Golowich:2007ka}.

\subsubsection{$c \to u \nu \bar{\nu}$  decay}
Within the SM, the $c \to u \nu \bar{\nu}$ transition is induced by box and $Z$-penguin diagrams. The corresponding short-distance effective Hamiltonian is given by \cite{Golowich:2009ii}
\begin{equation}
\mathcal{H}_{\rm eff}^{\rm SM} = \sum_{\ell=e,\mu,\tau} C_\ell^{\rm SM} \left(\bar{u} \gamma^{\mu} (1-\gamma_5) c \right) \left(\bar{\nu_\ell} \gamma_\mu (1-\gamma_5) \nu_\ell \right)\,,
\end{equation}
where
\begin{equation}
C_\ell^{\rm SM} = - \frac{G_F}{\sqrt{2}} \frac{\alpha_{\rm em}}{2 \pi \sin^2{\theta_W}} \sum_{q=d,s,b} \lambda_q X^\ell (x_q)\,.
\end{equation}
Here $\lambda_q = V_{uq} V_{cq} ^* $ and the structure functions are defined by $X^\ell(x_q) = \bar{D}(x_q,\,y_\ell)/2$, where $\bar{D}(x_q,\,y_\ell)$ is given as \cite{Inami:1980fz}
\begin{eqnarray}
\bar{D}(x_q,\,y_\ell) &=& \frac{1}{8} \frac{x_q \,y_\ell}{x_q - y_\ell} \left(\frac{y_\ell - 4}{y_\ell - 1}\right)^2 \ln y_\ell + \frac{x_q}{4} \nonumber\\ &&
+ \frac{x_q}{8} \left[\frac{x_q}{y_\ell -x_q}\left(\frac{x_q-4}{x_q-1}\right)^2 + 1 + \frac{3}{(x_q-1)^2}\right]\ln x_q \nonumber\\ &&  
- \frac{3}{8} \left(1+\frac{3}{y_\ell-1}\right)\frac{x_q}{x_q-1} \,,
\end{eqnarray}
with $x_q=m_q^2/M^2_W$ and $y_l=m_l^2/M^2_W$.

The $Z'$ contributes to $c \to u \nu \bar{\nu}$ transition through the seventh term in eq.~(\ref{Leff}). This contribution modifies the WC as $C_\ell \rightarrow C_\ell^{\rm SM}  + \mathcal{C}_{\mu}^{\nu\nu}$, where 
\begin{equation}
  \mathcal{C}_{\mu}^{\nu\nu} = \frac{h_L^{cu}g_L^{\mu \mu}}{4M_{Z'}^2}\,.
\end{equation}

 \section{Constraints on NP couplings}
\label{sec:III}
  In the literature there are several works where the rare charm decays have been 
 studied in the context of $Z'$ models. However, in most of these works, the constraints coming from $b \to s\, \mu^+ \, \mu^-$ sector were not considered. To emphasize the fact that $b \to s\, \mu^+ \, \mu^-$ sector can provide valuable constraints on the allowed new physics parameter space and hence restrict large new physics contributions to several observables in the up sector, we perform four types of fit:
 \begin{itemize}
\item F1: only $b \to s \nu \bar{\nu}$ data,
\item F2: $b \to s \nu \bar{\nu}$ data and $B_s - \bar{B}_s$ mixing,
\item F3: $b \to s \nu \bar{\nu}$ data, $B_s - \bar{B}_s$ mixing and neutrino trident,
\item F4: $b \to s \nu \bar{\nu}$ data, $B_s - \bar{B}_s$ mixing, neutrino trident and $b \to s\, \mu^+ \, \mu^-$ data. 
 \end{itemize}
 
\begin{table}[h!]
  \begin{center}
\begin{tabular}{|cc|cc|}
\toprule
Observable &  Value & Observable & Value \\ \hline
$G_F$	&  $1.16637 \times 10^{-5}$ \cite{pdg} & $B(B^0 \rightarrow K^0 \nu \bar \nu )$ & $2.6 \times 10^{-5}$ \cite{Grygier:2017tzo}\\ 
$\alpha_e (m_Z)$	& 1/127.9 \cite{pdg}& $B(B^0 \rightarrow K^{*0} \nu \bar \nu )$ & $1.8 \times 10^{-5}$ \cite{Grygier:2017tzo} \\ 
$\sin^2\theta_W$ &  0.23121 \cite{pdg}&  $B(B^+ \rightarrow K^+ \nu \bar \nu )$& $1.6 \times 10^{-5}$ \cite{Lees:2013kla}\\ 
$\lambda$& $0.22650\pm0.00048$ \cite{pdg}& $B(B^+ \rightarrow K^{*+} \nu \bar \nu )$  & $4.0 \times 10^{-5} $ \cite{Lutz:2013ftz}\\ 
$A$&  $0.790^{+0.017}_{-0.012}$ \cite{pdg}& $\Delta M_s^{\rm SM}/ \Delta M_s^{\rm Exp}$ & $1.04^{+0.04}_{-0.07}$ \cite{DiLuzio:2019jyq} \\ 
$\bar{\rho}$	&  $0.141^{+0.016}_{-0.017}$ \cite{pdg} & $R_{\nu}$ & $0.82 \pm 0.28$ \cite{Mishra:1991bv, Altmannshofer:2019zhy}\\ 
$\bar{\eta}$	& $0.357 \pm 0.011$ \cite{pdg}& $C_9^{\rm NP}$  & $-1.01 \pm 0.15$ \cite{Alok:2019ufo}\\  
$M_W$&  $80.385$ \cite{pdg}&  $C_9^{\rm NP}=-C_{10}^{\rm NP}$ & $-0.49 \pm 0.07$ \cite{Alok:2019ufo}\\ 
$m_t$&  $162.5$  \cite{pdg}&   &\\
 \hline
\end{tabular}
\caption{Various inputs used in the fits.}
\label{table-fit-input}
\end{center}
\end{table}

In the following, we discuss all observables used in the fit. These observables can be classified into following four sectors:

{\it $b \to s \nu \bar{\nu}$ sector:} The quark level transition $b \to s \nu \bar{\nu}$ induces exclusive semi-leptonic decays $B \to K^{(*)}\nu\bar{\nu}$. The experimental measurement of $b \to s \nu \bar{\nu}$ decay modes are challenging due to the presence of two final state neutrinos and require clean environment facilities such as  $e^+ e^-$ colliders.  At present, we only have upper bound in $b \to s \nu \bar{\nu}$ sector coming from Belle \cite{Grygier:2017tzo,Lutz:2013ftz} and BaBar \cite{Lees:2013kla} experiments. The current upper limits for $B \to K^{(*)}\nu\bar{\nu}$ decays at 90\% C.L. are \cite{Grygier:2017tzo,Lees:2013kla,Lutz:2013ftz}
\begin{eqnarray}
{B}(B^0 \rightarrow K^0 \nu \bar \nu ) &<& 2.6 \times 10^{-5} ~, \nonumber\\
{B}(B^0 \rightarrow K^{*0} \nu \bar \nu ) &<& 1.8 \times 10^{-5} ~,\nonumber\\
{B}(B^+ \rightarrow K^+ \nu \bar \nu ) &<& 1.6 \times 10^{-5} ~, \nonumber\\
{B}(B^+ \rightarrow K^{*+} \nu \bar \nu ) &<& 4.0 \times 10^{-5} ~.
\end{eqnarray}
These upper bounds are a factor of two to five above the SM predictions. With 50 $\rm ab^{-1}$ of data, Belle-II experiment is expected to observe these decay modes to an accuracy of 10\% on the branching ratio \cite{Belle-II:2018jsg,Bause:2021ply}. In order to include  these observables in the fit, we convert the above upper limits to a branching ratio as $\rm (0.5\, UB \pm 0.625\, UB)$ where $\rm UB$ is the experimental upper bound at 90\% C.L. The $\chi^2$  is then written as
\begin{eqnarray}
\chi^2_{b \to s \nu \bar{\nu}} &=&  \left(\frac{{B}(B^0 \rightarrow K^0 \nu \bar \nu )-1.3 \times 10^{-5}}{0.81 \times 10^{-5}}\right)^2 + \left(\frac{{B}(B^0 \rightarrow K^{*0} \nu \bar \nu )-0.9 \times 10^{-5}}{0.56 \times 10^{-5}}\right)^2 \nonumber\\
&&+ \left(\frac{{B}(B^+ \rightarrow K^+ \nu \bar \nu ) -0.8 \times 10^{-5}}{0.5 \times 10^{-5}}\right)^2 + \left(\frac{{B}(B^+ \rightarrow K^{*+} \nu \bar \nu )-2.0 \times 10^{-5}}{1.25 \times 10^{-5}}\right)^2\,,
\end{eqnarray}
where the theoretical predictions for $B \to K^{(*)}\nu\nu$ branching ratios are obtained using {\tt Flavio} \cite{Straub:2018kue}. We do not use constraints from the branching ratio of $B_s \to \phi \nu \bar{\nu}$ as it is much weaker than $B \to K^{(*)}\nu\bar{\nu}$ decay modes.

{\it $B_s - \bar{B}_s$ sector:} Here we consider constraints coming from the mass difference 
$\Delta M_s$. Using eq.~(\ref{eq:Bsmix}), the constraints on $g_L^{bs}$ can be translated in the following form
\begin{equation}
(g_L^{bs})^2 = (7.69 \pm 12.94) \times 10^{-6}\,.
\end{equation}
Here we used the value of $\Delta M_s^{\rm SM}/\Delta M_s^{\rm Exp}$ given in Table \ref{table-fit-input}. Hence the contribution of $\Delta M_s$ to $\chi^2$ is
\begin{equation}
\chi^2_{\Delta M_s} = \left(\frac{(g_L^{bs})^2- 7.69  \times 10^{-6}}{12.94 \times 10^{-6}}\right)^2\,.
\end{equation}

{\it Neutrino trident:} 
The contribution to the total $\chi^2$ coming from neutrino trident production is 
\begin{equation}
\chi^2_{\rm {trident}} = \left(\frac{R_\nu - 0.82}{0.28}\right)^2\,.
\end{equation}
The theoretical expression of $R_\nu$ is given in eq.~(\ref{trident}) whereas the experimental value is taken from Table \ref{table-fit-input}.

{\it $b \to s\, \mu^+ \, \mu^-$ sector:} A fit to all $b \to s\, \mu^+ \, \mu^-$ data leads to constraints on $C^{\rm NP}_9$  and $C^{\rm NP}_9 =-\, C^{\rm NP}_{10}$ scenarios as given in Table \ref{table-fit-input}. The fit value for $C^{\rm NP}_9$ ($C^{\rm NP}_9 = - \, C^{\rm NP}_{10})$ provides constraints on $Z_1$ ($Z_2$) model. Using the fit values and other inputs from Table I along with eq.~(\ref{bqllNP}), we get
\begin{eqnarray}
(g_L^{bs} g_L^{\mu\mu})_{Z_1} &=& (-8.36 \pm 1.24) \times 10^{-4}\,, \\
(g_L^{bs} g_L^{\mu\mu})_{Z_2} &=& (-8.11 \pm 1.16) \times 10^{-4} \,.
\end{eqnarray}

We use above constraints on $(g_L^{bs} g_L^{\mu\mu})$ in the fit. Hence the $\chi^2$ function can be written as 
\begin{eqnarray}
\chi^2_{b\to s \mu \mu, \, \rm Z_1}   &=& \left(\frac{(g_L^{bs} g_L^{\mu\mu}) - (-8.36 \times 10^{-4}) }{1.24 \times 10^{-4}}\right)^2\,, \\
\chi^2_{b\to s \mu \mu, \, \rm Z_2}  &=& \left(\frac{(g_L^{bs} g_L^{\mu\mu}) - (-8.11 \times 10^{-4} }{1.16 \times 10^{-4}}\right)^2\,.
\end{eqnarray}

Therefore, the $\chi^2$  for fits F1-F4 can be written as
\begin{eqnarray}
\chi^2_{\rm F1} &=& \chi^2_{b\to s \nu \bar{\nu}}  \, ,\\
\chi^2_{\rm F2} &=& \chi^2_{b\to s \nu \bar{\nu}} + \chi^2_{\Delta M_s}   \, ,\\
\chi^2_{\rm F3} &=& \chi^2_{b\to s \nu \bar{\nu}} + \chi^2_{\Delta M_s}   + \chi^2_{\rm trident}  \, ,\\
\chi^2_{\rm F4} &=& \chi^2_{b\to s \nu \bar{\nu}} + \chi^2_{\Delta M_s}   + \chi^2_{\rm trident} +  \chi^2_{b\to s \mu \mu} \,.
\end{eqnarray}
 The $\chi^2$ fit is performed using the CERN minimization code {\tt MINUIT} \cite{james}. 
The mass of $Z'$ is assumed to be 1 TeV in the fits.

\begin{figure}[ht]
\centering
\includegraphics[width = 3.2in]{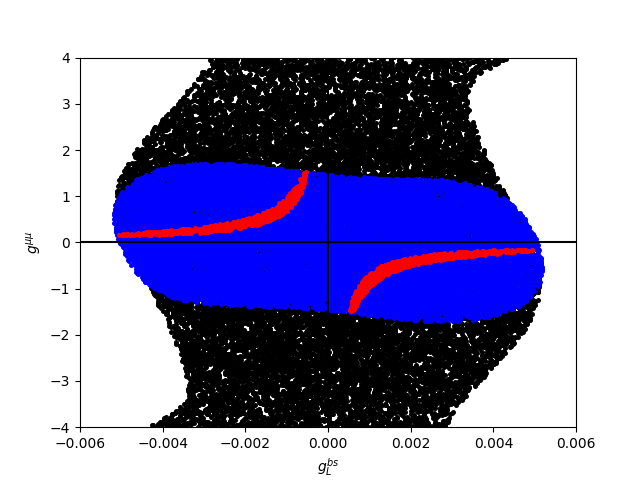}
\includegraphics[width = 3.2in]{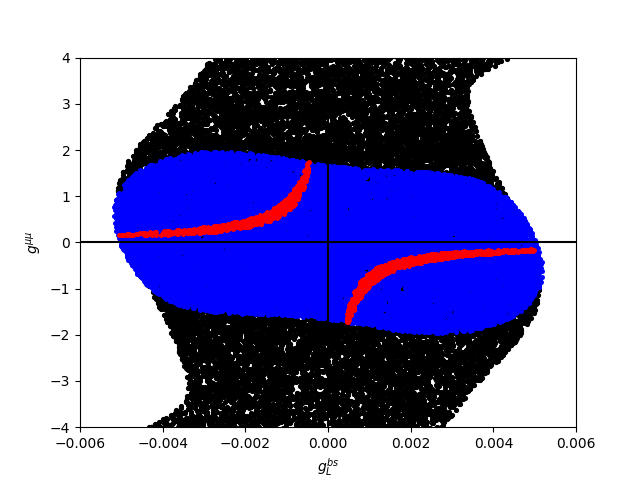}
\caption{The left and right panels depict 1$\sigma$-favoured parameter space of $(g_L^{bs},\, g_L^{\mu\mu})$  couplings for $Z_1$ and $Z_2$ models, respectively. The black, blue and red regions show allowed parameter space corresponding to $\rm F2$, $\rm F3$ and $\rm F4$ fits, respectively. The allowed regions correspond to $M_{Z'}$ =  1 TeV.}
\label{paramate1}
\end{figure}

For $\rm F1$ fit, $\chi^2$ is a function of the product $(g_L^{bs} g_L^{\mu\mu})$. The fit result is 
\begin{equation}
(g_L^{bs} g_L^{\mu\mu}) = -0.0047 \pm 0.0051.
\end{equation}
 The results for $\rm F2$, $\rm F3$ and $\rm F4$ fits for $Z_1$ and $Z_2$ models are presented in Fig.~\ref{paramate1}. It is obvious from the figure that if we only include constraints coming from  $b \to s \nu \bar{\nu}$   and $\Delta M_s$, a relatively large parameter space is allowed. Therefore while correlating $c \to u $ decays, in particular $c \to u \nu \bar{\nu}$,  with $b \to s$ sector if only these constraints are considered, a possibility of large new physics effects may survive.

However, the parameter space shrinks after including neutrino trident in the fit. Finally, after including $b \to s \mu^+ \mu^-$ constraints,  the allowed range of new physics couplings reduces considerably. Therefore it will be interesting to see whether any useful new physics effects in charm sector, particularly  decays induced by $c \to u \nu \bar{\nu}$ and $c \to u \mu^+ \mu^-$ transitions, is allowed by the current $b \to s $ and neutrino trident data.
  
 \section{Semi-leptonic charm decays and mixing}
 \label{sec:IV}
 Rare or forbidden decays of charm mesons provide a unique channel to probe new physics in the up-type quark sector \cite{Gisbert:2020vjx}. This is complementary to new physics searches in the first and third generations of quarks as well as in the down-quark sectors. The fact that the charm quark is not as heavy as bottom, the application of heavy quark effective theory is not as useful as it is for the B-meson sector. Also the mass of charm quark is not small enough to be considered as a light quark. Therefore predictions in the charm sector are usually dominated by the long-distance effects. Within the SM, the  short-distance contributions to the purely leptonic and semi-leptonic charm decays along with $D-\bar{D}$ mixing are highly suppressed, in fact almost negligible, owing to GIM cancellation and CKM suppression. Probing new physics then becomes a challenging task as an unambiguous signal would require enhancements above the level of long-distance contributions. 
 
 In the following subsections we discuss the impact of measurements in B meson sector on 
 rare FCNC processes induced by the $c \to u $ transition in the context of non-universal $Z'$ models. We consider decays induced by  $c \to u  \nu \bar{\nu}$ and $c \to u  \mu^+ \mu^-$ transitions along with $D-\bar{D}$ mixing. The primary motive is to identify observable(s) for which the current $b \to s$ data allows short-distance  contributions above the long-distance contributions.

\subsection{Branching ratio of dineutrino charm decays}
The new physics searches can be exquisitely performed using the null-tests of the SM. In the up-sector, $c \to u$  transitions provide such null tests due to strong CKM suppression. In this regard, the decays induced by the quark level  transition $c\to u \nu \bar{\nu}$ are particularly interesting as the long-distance effects are relatively smaller in comparison to
their  charged dileptons counterparts. Hence in $c \to u$  sector, dineutrino  modes can be considered to provide genuine null-tests of SM \cite{Gisbert:2020vjx,Bause:2020auq,Bause:2020xzj,Colangelo:2021myn,Faisel:2020php}. The $c\to u \nu \bar{\nu}$ transition induces several exclusive decay modes such as $D^+ \to \pi^+ \nu \bar{\nu}$, $D^0\to \pi^0 \nu \bar{\nu}$ and $B_c^+ \to B^+ \nu \bar{\nu}$. Any observation of these decays at the level of current experimental sensitivity can be considered as unambiguous signature of beyond SM physics. 

On the experimental front, currently there are no experimental upper limits on any of the semi-leptonic decays included by $c\to u \nu \bar{\nu}$ transition. We only have upper limits 
obtained using the  $SU(2)_L$ invariance and bounds on the charged lepton decay modes which allow branching ratios to be as high as a few times $10^{-5}$ \cite{Bause:2020auq}. These limits can go down if charged leptons bounds are improved. Owing to the clean hadronic environements, the $e^-e^-$ colliders such as Belle II \cite{Belle-II:2018jsg}, BES III \cite{BESIII:2020nme} as well as the future colliders, for instance FCC-ee \cite{FCC:2018byv} are well suited for the studies related to the dineurino  decay modes.  For efficiencies of a permille or better, $D$ decay modes can be observed at the Belle II and FCC-ee experiments provided new physics enhances their branching ratios up to a level of $O(10^{-6})$ - $O(10^{-8})$ \cite{Bause:2020xzj}.

In this section we analyse $D^0 \to \pi^0 \nu \bar{\nu}$, $D^+ \to \pi^+ \nu \bar{\nu}$ and $B_c^+ \to B^+ \nu \bar{\nu}$  decays  in  the context of non-universal $Z'$ models.
The differential branching ratio of  $M_1 \to M_2 \nu \bar{\nu}$ mode, where $M_1=D^0,D^+,B_c$ and $M_2=\pi^0,\pi^+$ and $B^+$, respectively is given by
\begin{equation}
\frac{dB}{dq^2}= \frac{\tau_{M_1} \,\left|f_+(q^2)\right|^2 \,\lambda^{3/2}}{32 \,\pi^3\, m^3_{M_1}} |C_{\ell}|^2,
\end{equation}
where $q^2$ denotes the invariant mass-squared of the dineutrinos, $\lambda\equiv \lambda(m^2_{M_1}, m^2_{M_2}, q^2)$ is the Kallen function defined as $\lambda(x,y,z)=(x + y +z)^2 -4(xy+yz+zx)$ and $\tau_{M_1}$ is the lifetime of the $M_1$ meson. The $M_1 \to M_2$ form-factors are defined as
\begin{eqnarray}
\langle M_2(p')|\bar u \gamma_{\mu}c|M_1(p)\rangle
  &=& f_+ (q^2) \left( p_{\mu} + p'_{\mu} -\frac{m^2_{M_1}-m^2_{M_2}}{q^2}q_{\mu}\right)  + f_0(q^2) \frac{m^2_{M_1}-m^2_{M_2}}{q^2}q_{\mu}\,.
   \label{ff-def}
\end{eqnarray}
The $D \to \pi$ and $B_c \to B$ form-factors are calculated within the framework of lattice QCD in refs. \cite{Lubicz:2017syv} and \cite{Cooper:2020wnj}, respectively.

The integrated branching ratio is then given by
\begin{equation}
B(M_1 \to M_2 \nu \bar{\nu}) = \frac{\tau_{M_1}\,|C_{\ell}|^2}{32 \,\pi^3\, m^3_{M_1}} \int_{q^2_{\rm min}}^{q^2_{\rm max}} \lambda^{3/2} \left|f_+(q^2)\right|^2\,dq^2\,.
\end{equation}
For $D^0 \to \pi^0 \nu \bar{\nu}$ mode, $q^2_{\rm min}=0$ whereas $q^2_{\rm max}=(m_{D^0}-m_{\pi^0})^2$. In $D^+ \to \pi^+ \nu \bar{\nu}$ decay, resonant backgrounds through $\tau$-leptons lead to the same final state. This is removed by phase space cuts due to which $q^2_{\rm min}$ for $D^+ \to \pi^+ \nu \bar{\nu}$  mode is $(m^2_{\tau}-m^2_{\pi^+})(m^2_{D^+}-m^2_{\tau})/m^2_{\tau}\approx 0.34$ $\rm GeV^2$. The $q^2_{\rm max}$ for this mode is $(m_{D^+}-m_{\pi^+})^2$. For $B_c \to B^+ \nu \bar{\nu}$ mode, $q^2_{\rm min}=0$ whereas $q^2_{\rm max}=(m_{B_c}-m_{B^+})^2$. Using the  form-factor  calculations in \cite{Lubicz:2017syv,Cooper:2020wnj}, we get
\begin{eqnarray}
B(D^0 \to \pi^0 \nu \bar{\nu}) &\approx & 1.0 \times 10^9  \, |C_{\ell}|^2, \\
B(D^+ \to \pi^+ \nu \bar{\nu}) &\approx & 4.1 \times 10^9   \, |C_{\ell}|^2, \\
B(B_c \to B^+ \nu \bar{\nu})   &\approx & 7.7 \times 10^8    \, |C_{\ell}|^2.
\end{eqnarray}

The short-distance SM predictions for the branching ratios of $D^0 \to \pi^0 \nu \bar{\nu}$ and $D^+ \to \pi^+ \nu \bar{\nu}$ are $\sim 2 \times 10^{-16}$ and $\sim 8 \times 10^{-16}$, respectively. The long-distance contributions in these decay modes are predicted to be $\sim 10^{-16}$ \cite{Burdman:2001tf}. For $B_c \to B^+ \nu \bar{\nu}$ mode, the short-distance SM contribution is $\sim 1.5 \times 10^{-16}$ whereas the long-distance contribution is estimated to be $\sim 10^{-16}$ \cite{Colangelo:2021myn}. Therefore for unambiguous signature of new physics in these decay modes, the short-distance branching ratio should be enhanced well above $10^{-16}$. 

 The decay mode  $D^+ \to \pi^+ \nu \bar{\nu}$ in $Z'$ models was studied in \cite{Bause:2020xzj}. It was shown that the branching ratio of $D^+ \to \pi^+ \nu \bar{\nu}$ can be enhanced up to $\sim 10^{-6}$ which falls in the same ballpark as the model-independent upper bound.  In \cite{Colangelo:2021myn}, it was shown that the branching ratio of $B_c^+ \to B^+ \nu \bar{\nu}$ in the 331 model can be enhanced up to $\sim 10^{-11}$. It would be interesting to see whether such large enhancements is allowed by the current $b \to s$ and neutrino trident data in the non universal  $Z'$ model considered in this work.

\begin{table}[h]
\centering
\tabcolsep 2pt
\begin{tabular}{|c|c|c|c|c|c|}
\hline\hline
Constraints  & $C_{\ell}^{\rm max}$ & $B(D^0 \to \pi^0 \nu \bar{\nu})$ & $B(D^+ \to \pi^+ \nu \bar{\nu})$ & $B(B_c \to B^+ \nu \bar{\nu})$ \\
\hline
$b \to s \nu \nu$  & $7.4 \times 10^{-11}$   & $5.6 \times 10^{-12}$ & $2.2 \times 10^{-11}$ &  $4.2 \times 10^{-12} $ \\
\hline
$b \to s \nu \nu$, $\Delta M_s$  & $7.4 \times 10^{-11}$  &  $5.6 \times 10^{-12}$ & $2.2 \times 10^{-11}$  &  $4.3 \times 10^{-12}$\\
\hline
$b \to s \nu \nu$, $\Delta M_s$ , Neutrino trident  & $1.4 \times 10^{-11}$   & $2.1 \times 10^{-13}$ &  $8.8 \times 10^{-13}$  &  $1.6 \times 10^{-13}$ \\
\hline
$b \to s \nu \nu$, $\Delta M_s$ , Neutrino trident, $b \to s \ell^+ \ell^-$   & $2.3 \times 10^{-12}$   & $5.5 \times 10^{-15}$ & $2.2 \times 10^{-14}$ & $4.2 \times 10^{-15}$ \\

\hline\hline
\end{tabular}
\caption{1$\sigma$ upper limit on the branching ratios obtained by using constraints on $Z'$ couplings for $Z_1$ model.}
\label{tab:pred}
\end{table}

\begin{figure}[ht]
\centering
\includegraphics[width = 2.1in]{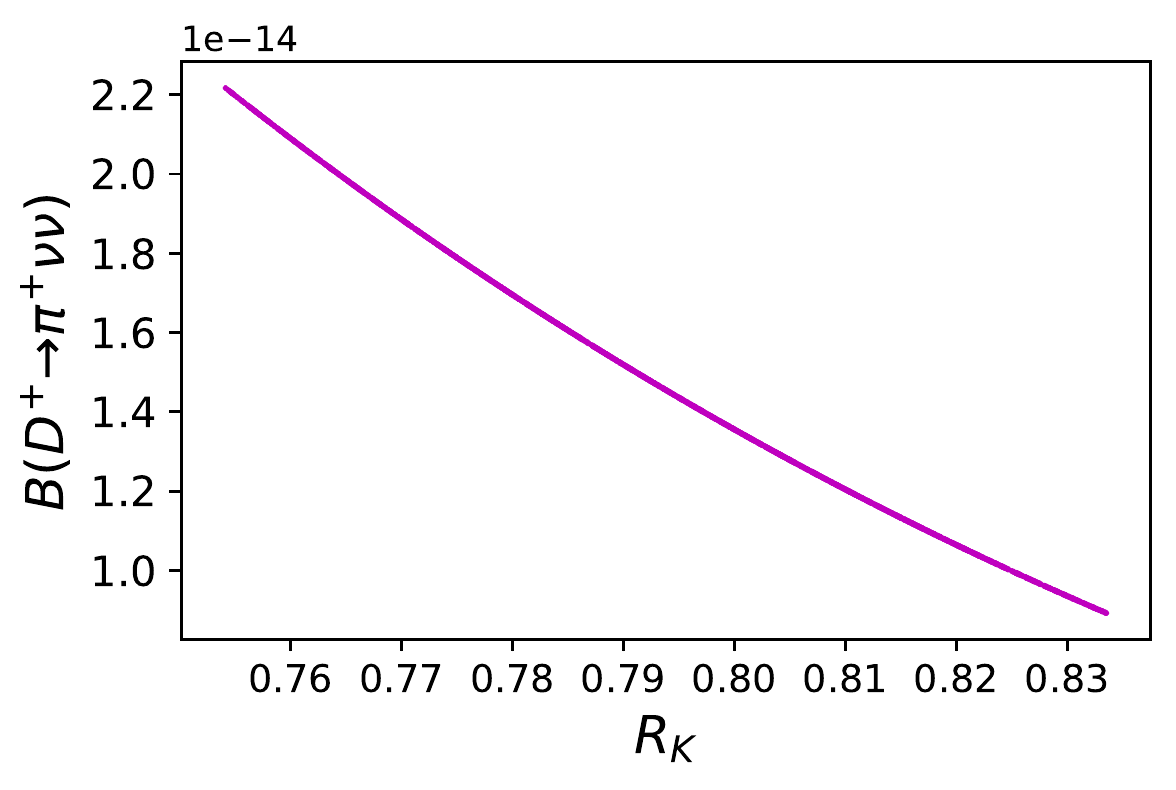}
\includegraphics[width = 2.1in]{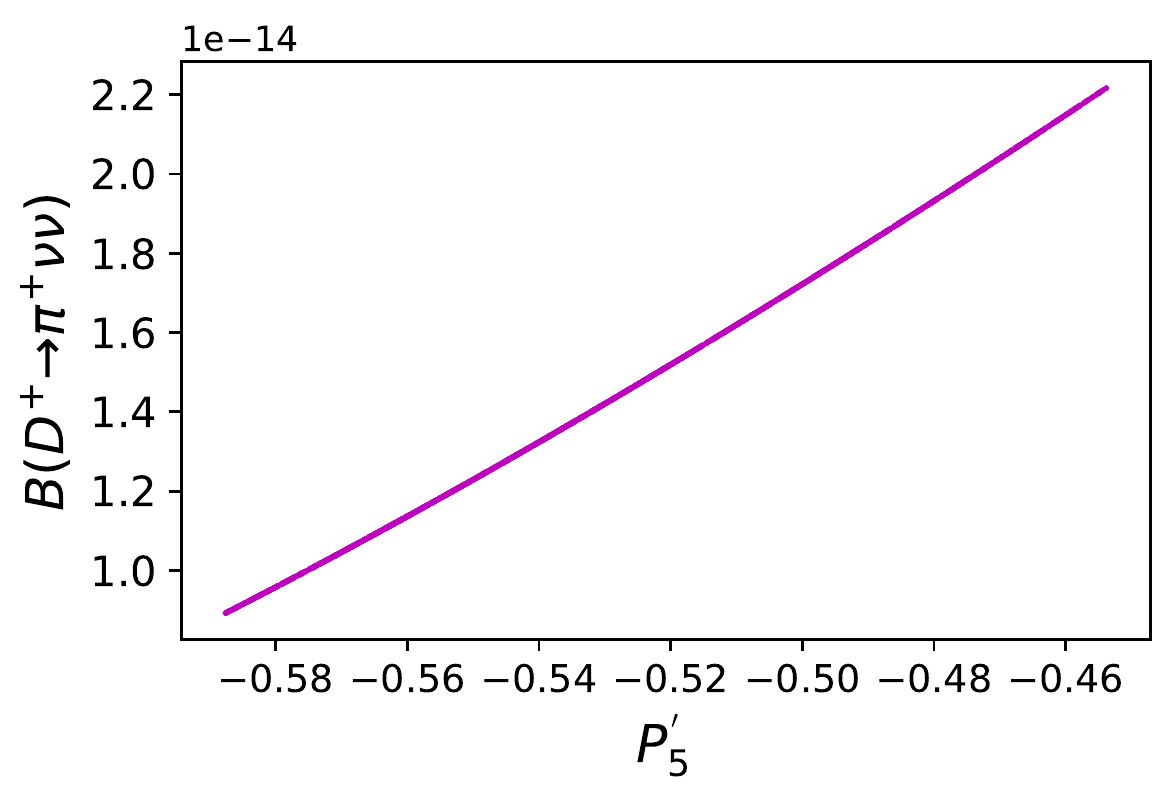}
\includegraphics[width = 2.1in]{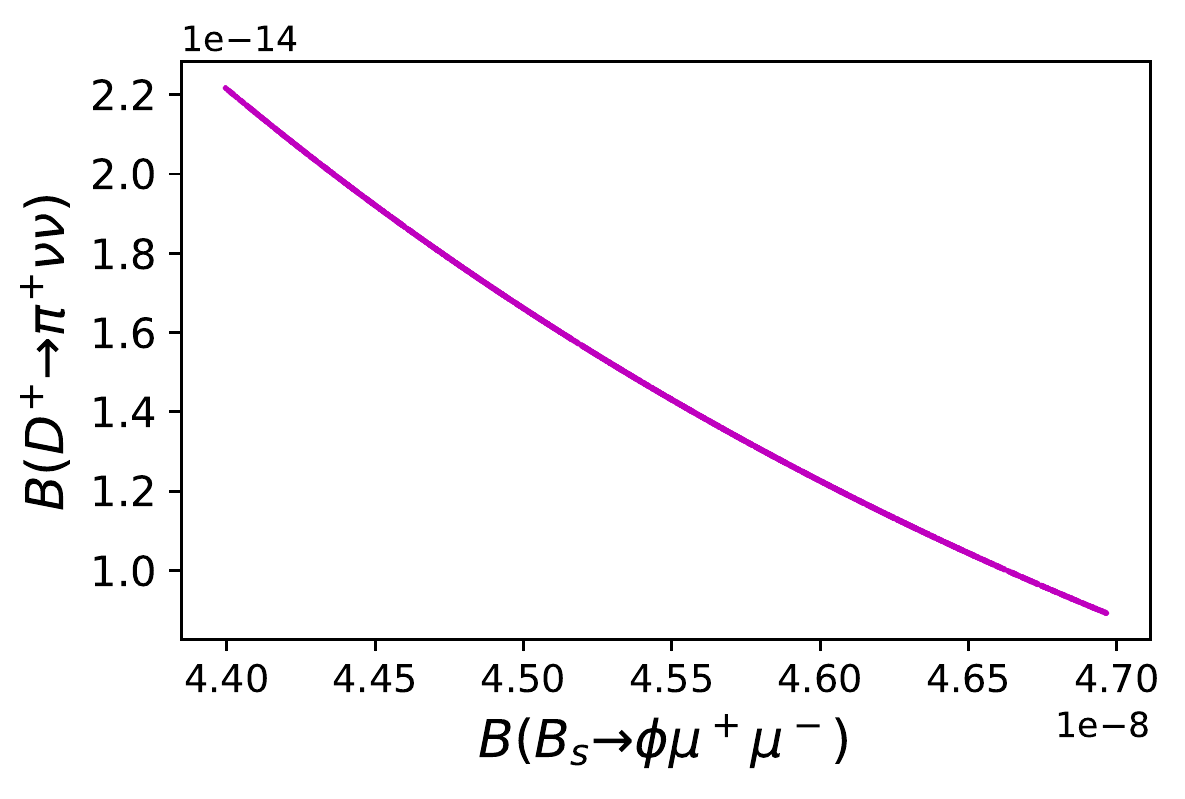}
\caption{Plots indicating correlations between the branching ratio of $D^+ \to \pi^+ \nu \bar{\nu}$ and $b \to s \mu^+ \mu^-$ observables $R_K$,  $P'_5$ and $B(B_s \to {\phi} \mu^+ \mu^-)$  in $Z_1$ model. The correlations are almost the same for $Z_2$ model.}
\label{fig:corrbsll}
\end{figure}

Using the allowed range of $Z'$ couplings obtained in Sec.~\ref{sec:III}, the 1$\sigma$ upper limit on $B(D^0 \to \pi^0 \nu \bar{\nu})$, $B(D^+ \to \pi^+ \nu \bar{\nu})$ and $B(B_c \to B^+ \nu \bar{\nu})$ for $Z_1$ model are given in Table \ref{tab:pred}. 
It is obvious that if we only use constraints from dineutrino bottom decays, about four orders of magnitude enhancement in the branching ratio is allowed. The conclusions remain almost the same if $\Delta M_s$ is included in the fit along with $b \to s \nu \bar{\nu}$.  However there is a large reduction in the allowed parameter  space if constraints from neutrino trident are included. The parameter space shrinks  further by additional of  $b \to s \ell^+ \ell^-$ data. For the combined fit F4, the branching ratios can only be enhanced up to an order of magnitude above the SM value. This is also evident from Fig.~\ref{fig:corrbsll} which illustrates correlations between $B(D^+ \to \pi^+ \nu \bar{\nu})$ and $b \to s \mu^+ \mu^-$ observables $R_K$,  $P'_5$ and $B(B_s \to {\phi} \mu^+ \mu^-)$ in $Z_1$ model. The conclusions are almost the same for the $Z_2$ model. This enhancement is not useful in the sense that it is orders of magnitude smaller than the detection level for these modes in any of the currently running  or planned collider facilities.

\subsection{Branching ratio of dimuon charm decays}
In this section we consider $D^0 \to \mu^+ \mu^-$  and $D^+ \to \pi^+ \mu^+ \mu^-$ decays  which are induced by the quark level transition $c \to u \mu^+ \mu^-$. In the SM, the short-distance contributions to these decay modes  are highly suppressed and the dominant contribution comes from   the long-distance effects. It would be interesting to see whether the new physics contributions coming from the addition of $Z'$ can enhance the decay rates above the long-distance contributions.

\subsubsection{$D^0 \to \mu^+ \mu^-$ decay}
The short-distance contribution to the branching ratio of $D^0 \to \mu^+ \mu^-$ decay is negligible in the SM. This is due to two fold effect:
\begin{itemize}
\item Within SM, only $C_{10}^{cu}$ WC contributes and its value is negligibly small.
\item $D^0 \to \mu^+ \mu^-$ suffers from helicity suppression.
\end{itemize}
In the SM,  the branching ratio of $D^0 \to \mu^+ \mu^-$ decay is dominated by the long-distance effects in $D^0 \to \gamma^* \gamma^*  \to \mu^+ \mu^-$ \cite{Burdman:2001tf,Petrov:2016kmb,Petrov:2017nwo}. Using the upper bound on $D^0 \to \gamma \gamma < 8.5 \times 10^{-7}$   at 90\% C.L. \cite{pdg}, the SM branching ratio  is estimated to be \cite{Fajfer:2015mia,Gisbert:2020vjx}
\begin{eqnarray}
{B}(D^0 \to \mu^+ \mu^-) _{\rm LD} &\approx & 8\alpha^2\, \left(\frac{m_{\mu}^2}{m_{D^0}^2}\right)\, \log^2 \left(\frac{m_{\mu}^2}{m_{D^0}^2}\right) \times {B}(D^0 \to \gamma \gamma)  \sim 10^{-11}.
\end{eqnarray}
From the experimental side, we only have an upper bound. The most stringent upper bound is
reported by the LHCb experiment in 2013 which is $6.2 \times 10^{-9}$ at 90\% C.L.
using the data-set corresponding to an integrated luminosity of  0.9 $\rm fb^{-1}$ \cite{Aaij:2013cza}. Therefore the SM value of the branching ratio including the long-distance effects is well below the current experimental upper limit. Hence one can safely neglect the SM contribution while calculating  the branching ration in $Z'$ model. Under this assumption,  the branching ratio is given by
\begin{eqnarray}
{B}(D^0 \to \mu^+ \mu^-)_{Z'} &=& \frac{\tau_{D^0} f_D^2 m_{\mu}^2 m_{D^0}}{32 \pi M^4_{Z'}}\sqrt{1-\frac{4m_{\mu}^2}{m_{D^0}^2}}  \times (h^{cu}_L)^2 (g_L^{\mu\mu}-g_R^{\mu\mu})^2,
\end{eqnarray}
where $f_D = 212.0 \pm 0.7$ MeV \cite{FlavourLatticeAveragingGroup:2019iem} and $\tau_{D^0}= 4.101 \times 10^{-13}$ s \cite{pdg}. It should be noted that ${B}(D^0 \to \mu^+ \mu^-)_{Z'}=0$ for $Z_1$ model.

Using constraints obtained on new physics couplings for $Z_2$ model, we find that ${B}(D^0 \to \mu^+ \mu^-)_{Z'} $ can be as large as $5 \times 10^{-13}$ for the F1 fit. For fits F2, F3 and F4,  the upper bounds on  ${B}(D^0 \to \mu^+ \mu^-)_{Z'} $ are $5 \times 10^{-13}$,   $2 \times 10^{-14}$ and  $5 \times 10^{-16}$, respectively. Thus it becomes apparent that the current $b \to s $ and neutrino trident data rules out any useful  enhancement in ${B}(D^0 \to \mu^+ \mu^-)_{Z'} $.

\subsubsection{$D^+ \to \pi^+ \mu^+ \mu^-$ decay}
The quark level transition $c \to u \mu^+ \mu^-$ generates semi-leptonic decay $D^+ \to \pi^+ \mu^+ \mu^-$. Within SM, the short-distance contribution to the total branching ratio is highly suppressed, $O(10^{-12})$ \cite{Fajfer:2015mia}. The dominant contribution comes from  vector resonances $\rho$, $\omega$ and $\phi$, decaying to dimuon pair. The effects of these resonances are incorporated assuming naive factorization by addition of $q^2$-dependent terms to $C_9^{cu}$ \cite{Fajfer:2015mia}. The contribution to the total branching ratio coming from vector resonances is $O(10^{-10})$ \cite{Sahoo:2017lzi}. On the experimental front, at present we only have upper bounds. In 2013, LHCb reported following upper limits at 90\% C.L. in the low- and high-$q^2$ bins using data corresponding to an integrated luminosity of 1.0 $\rm fb^{-1}$  \cite{LHCb:2013hxr}
\begin{eqnarray}
B (D^+ \to \pi^+ \mu^+ \mu^-)_{\rm low-q^2} &<& 2.0 \times 10^{-8}\,,\\
B (D^+ \to \pi^+ \mu^+ \mu^-)_{\rm high-q^2} &<& 2.6 \times 10^{-8}\,.
\end{eqnarray}
Here low- and high-$q^2$ bins correspond to $(0.0625,\, 0.276)$ $\rm GeV^{2}$ and $(1.56,\,4.00)$ $\rm GeV^{2}$, respectively. 

Given the current experimental accuracy, we neglect the SM contribution. Under this assumption, the differential branching ratio in the $Z'$ model, which generates $c \to u \mu^+ \mu^-$ transition through fifth term in eq.~(\ref{Leff}), is given by
\begin{eqnarray}
\left(\frac{dB}{dq^2}\right)_{Z'} &=& \frac{\tau_{D^+}\lambda^{1/2} \, \beta_{\mu}}{1024 \,\pi^3 \,m_{D^+}^3\,M^4_{Z'}}\,  \Bigg[a_1(q^2) (h^{cu}_L)^2 (g_L^{\mu\mu}+g_R^{\mu\mu})^2\,  +\, a_2(q^2) (h^{cu}_L)^2 (g_L^{\mu\mu}-g_R^{\mu\mu})^2 \Bigg]\,,
\end{eqnarray}
where $\lambda$ is the Kallen function, $\beta_{\mu}=\sqrt{1-4m_{\mu}^2/q^2}$ and $\tau_{D^+}= 1.04 \times 10^{-12}$ s \cite{pdg}. The functions $a_{1(2)}(q^2) $ are defined as
\begin{eqnarray}
a_1(q^2) &=& \left(\frac{\lambda}{2} - \frac{\lambda \beta_{\mu}^2}{6}\right) f^2_+(q^2)  \,,\\
a_2(q^2) &=& a_1(q^2) + 8 m_{\mu}^2 m_{D^+}^2 f^2_+(q^2)  + 2 q^2 m_{\mu}^2 F^2(q^2) - 4 m_{\mu}^2 \left(m_{D^+}^2 - m_{\pi^+}^2 + q^2 \right) f_+(q^2) F(q^2),
\end{eqnarray}
with 
\begin{equation}
F(q^2) = f_+(q^2)  - \frac{m_{D^+}^2 - m_{\pi^+}^2}{q^2 } \left(f_0(q^2)  -  f_+(q^2) \right)\,.
\end{equation}
The $D^+ \to \pi^+$ form-factors $f_+(q^2)$ and $f_0(q^2)$ are defined in eq.~(\ref{ff-def}). 
The $q^2$ dependence of these form-factors are given in the Becirevic-Kaidalov (BK) parametrization as \cite{Becirevic:1999kt}
\begin{eqnarray}
f_+(q^2) &=& \frac{f_+(0)}{(1-x)(1-ax)}\,,\\
f_0(q^2) &=& \frac{f_+(0)}{(1-x/b)}\,,
\end{eqnarray}
where $x\equiv q^2/m^2_{\rm pole}$ with $m_{\rm pole}=1.90 \pm 0.08$ GeV, $a=0.28 \pm 0.14$ and $b=1.27 \pm 0.17$ are the shape parameters \cite{Fajfer:2015mia}. These shape parameters are determined  by measurements of  $D \to \pi l \nu$ decay spectra. The form-factor $f_+(0)$  calculated by the HPQCD collaboration is $0.67 \pm 0.03$ \cite{Na:2011mc}.

For F1, F2, F3 and F4 fits, the $Z'$ contribution in $Z_1$ model to the branching ratio of $D^+ \to \pi^+ \mu^+ \mu^-$, $B(D^+ \to \pi^+ \mu^+ \mu^-)_{Z'}$, in the low-$q^2$ region  are $3 \times 10^{-12}$, $3 \times 10^{-12}$, $6 \times 10^{-14}$ and $3 \times 10^{-15}$, respectively.  The SM short distance prediction for $B(D^+ \to \pi^+ \mu^+ \mu^-)$ in the low-$q^2$ region is $3 \times 10^{-13}$. Thus we see that if we use constraints coming from the combined fit,  the new physics contribution is negligible.  The same is true in the high-$q^2$ region. The conclusion remains unchanged for the $Z_2$ model.

\subsection{Charm Mixing}

\begin{figure}[h!]
\centering
\includegraphics[width = 3.4in]{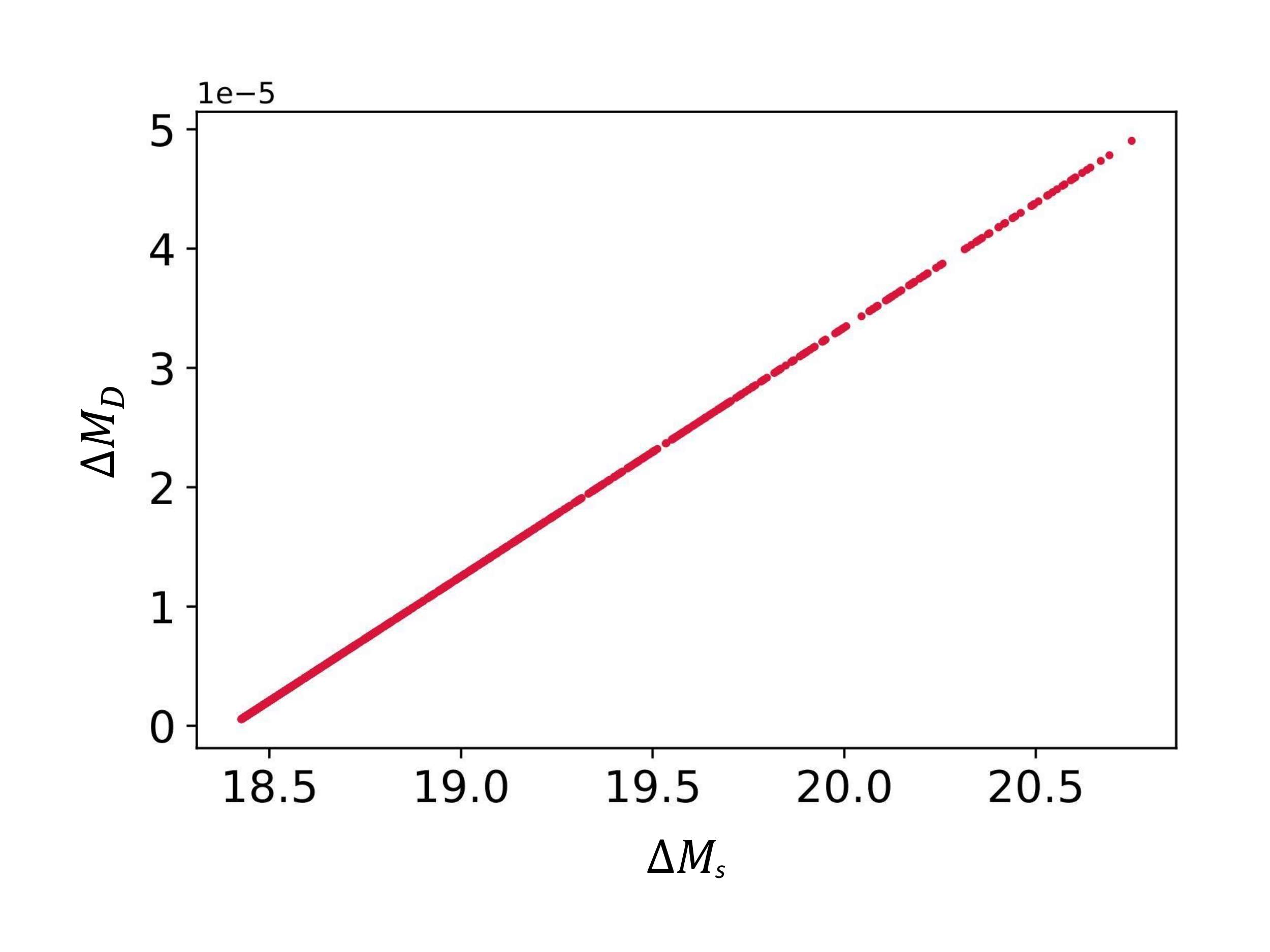}
\caption{Plot indicating correlations between $\Delta M_s$ and $\Delta M_D$. The values are in the units of $\rm ps^{-1}$.}
\label{fig:mix}
\end{figure}

The observed value of $\Delta M_D$ is much higher than the short-distance SM value but it is in qualitative accord with the long distance SM predictions. However as no reliable estimate of the later is obtained so far, the possibility  of large new physics contributions to $D-\bar{D}$ mixing is not ruled out. Here we consider such an effect in the context of $Z'$ model. In the current framework, the contribution to  $\Delta M_D$ coming from a $Z'$ boson is governed by $Z'bs$ coupling along with the CKM factor $(V_{us} V_{cb}^{*})$. Using the allowed range of $g_L^{bs}$ coupling from the current $b \to s$ data, we find that $(\Delta M_D)_{Z'} <  5 \times 10^{-5}\, {\rm ps^{-1}}$. A correlation between $\Delta M_D$ and $\Delta M_{s}$  is shown in Fig.~\ref{fig:mix}. Thus we see that a possibility of large new physics contributions to  $\Delta M_D$ coming from a $Z'$ boson is disfavoured by the current measurements in the $b \to s$ sector.

\section{Conclusions}
\label{sec:VI}
In this paper we consider  a non-universal $Z'$ model which generates $b\rightarrow s\, \mu^+\,\mu^-$ transition at the tree level. Considering  the down-type quarks in the quark-doublets to be in the mass-flavour diagonal basis, $u_i \to u_j$ transitions are induced by the up-type quarks mediated by $Z'bs$ and $Z'\mu\mu$ couplings along with suitable combinations of the elements of quark mixing matrix. Thus the current $b \to s$ data is expected to have potential impact on new physics contributions to several observables in the charm  sector. Here we study  such an impact on decays induced by $c \to u \nu\,\bar{\nu} $ and $c \to u \mu^+\,\mu^- $  transitions along with $\Delta M_D$. Performing a combined fit to $b\rightarrow s\, \mu^+\,\mu^-$, $\Delta M_s$ and neutrino trident data, we obtain following results:

\begin{itemize}

\item The SM predictions of $D^0 \to \pi^0 \nu \bar{\nu}$, $D^+ \to \pi^+ \nu \bar{\nu}$ and $B_c^+ \to B^+ \nu \bar{\nu}$  decays, including the long-distance contributions, are $\sim 10^{-16}$. We find that in $Z'$ models, the branching ratios of these decay  can be enhanced by an order of magnitude above their SM values. However this enhancement is not enough to enable the observation of these decay modes in any of the currently running and planned experimental facilities. 

\item The short distance contribution to the branching ratios of $D^0 \to \mu^+ \mu^-$  and $D^+ \to \pi^+ \mu^+ \mu^-$ are highly suppressed and within SM, these decay modes are dominated by the long-distance effects. The SM branching ratios of $D^0 \to \mu^+ \mu^-$  and $D^+ \to \pi^+ \mu^+ \mu^-$ are $\sim 10^{-11}$ and $\sim 10^{-10}$, respectively. The $Z'$ contribution to the branching ratios of  these decays are $\sim 10^{-16}$ and $\sim 10^{-15}$, respectively.

\item Within SM, the mixing observable $\Delta M_D$ is dominated by long-distance contributions. We find that the $Z'$ boson contribution is restricted to $\sim 10^{-5}\, {\rm ps^{-1}}$ which is well below the experimental value.
\end{itemize}

Thus we conclude that  the current data in the $B$ sector puts severe constraints on new physics contributions in semi-leptonic charm decays and mixing. Any useful enhancement in $c \to u \nu \bar{\nu}$, $c \to u \mu^+ \mu^-$ decays and $\Delta M_D$ is ruled out  in the context of $Z'$ model considered  in this work.

\bigskip
\noindent
{\bf Acknowledgements}: The work of A.K.A. is supported by SERB-India Grant CRG/2020/004576.
We would like to thank Shireen Gangal for useful discussions.

\end{document}